# Data Provenance Tracking as the Basis for a Biomedical Virtual Research Environment

**Richard McClatchey**[1]
*Centre for Complex Cooperative Systems, University of the West of England*
*Coldharbour Lane, Frenchay, Bristol BS16 1Qy  United Kingdom*
*E-mail:* `richard.mcclatchey@uwe.ac.uk`

**Abstract**: In complex data analyses it is increasingly important to capture information about the usage of data sets in addition to their preservation over time to ensure reproducibility of results, to verify the work of others and to ensure appropriate conditions data have been used for specific analyses. Scientific workflow based studies are beginning to realize the benefit of capturing this *provenance* of data and the activities used to process, transform and carry out studies on those data. This is especially true in biomedicine where the collection of data through experiment is costly and/or difficult to reproduce and where that data needs to be preserved over time. One way to support the development of workflows and their use in (collaborative) biomedical analyses is through the use of a Virtual Research Environment. The dynamic and distributed nature of Grid/Cloud computing, however, makes the capture and processing of provenance information a major research challenge. Furthermore most workflow provenance management services are designed only for data-flow oriented workflows and researchers are now realising that tracking data or workflows alone or separately is insufficient to support the scientific process. What is required for collaborative research is traceable and reproducible provenance support in a full orchestrated Virtual Research Environment (VRE) that enables researchers to define their studies in terms of the datasets and processes used, to monitor and visualize the outcome of their analyses and to log their results so that others users can call upon that acquired knowledge to support subsequent studies. We have extended the work carried out in the neuGRID and N4U projects in providing a so-called Virtual Laboratory to provide the foundation for a generic VRE in which sets of biomedical data (images, laboratory test results, patient records, epidemiological analyses etc.) and the workflows (pipelines) used to process those data, together with their provenance data and results sets are captured in the CRISTAL software. This paper outlines the functionality provided for a VRE by the Open Source CRISTAL software and examines how that can provide the foundations for a practice-based knowledge base for biomedicine and, potentially, for a wider research community.

---

[1]Professor Richard McClatchey, UWE, Bristol UK

1. Introduction

In the big data age, biomedical informatics systems need to be increasingly flexible, accessible and shareable in order to provide a usable platform for collaborative data analytics. Consequently, we need to design such systems to enable users to easily identify data and algorithms for carrying out analyses; the systems should capture 'provenance' information about how, when, by whom and why the system has been used. This facilitates reproducibility and verification of results, validation of analyses between researchers and traceability of analysis procedures over time as required in a Virtual Research Environment (or VRE [1]). Allowing systems to be self-describing is one way to enable this; there have been recent important advances recently which allow us to start to build genuinely self-describing systems based on the concepts of meta-data and meta-models, item description and, eventually, ontologies [2]. With the Cloud we are now able to manage large medical data sets from imaging, epidemiology and complex analyses and to provide the infrastructure on which clinical analytics and research can take place. This paper proposes a self-describing or description-driven approach to medical systems design; the exemplar of this approach is the CRISTAL system [3]. We show that this approach allows data structure definitions to be stored and versioned, leading to increased traceability, which facilitates longitudinal analyses in a VRE. The collected, traceability information includes not only data on the use of system Items (e.g. data, process, document or agent Items) but also how they have changed over time, by whom and for what purpose, providing a fine granularity in the provenance of the Items over the lifecycle of the system. We also demonstrate the application of these principles in a Virtual Laboratory and how this can form the basis of a VRE.

To date provenance gathering systems and techniques have mostly been used within scientific research domains such as neuroscience [3] or in bioinformatics [4]. Often provenance is gathered only as a result of the execution of scientific workflows [4] whereas we propose its active capture and management over both data and process evolution in a VRE. The usefulness of provenance data has been discussed at length elsewhere; the reader is directed to other important works such as [5]. There exists a large body of work on (distributed) workflow systems (e.g. [6, 7, 8, 9]) but to date none handle data and workflows uniformly to provide full system lifecycle provenance. We have developed an application of provenance management for a biomedical VRE, as discussed in this paper. The next section outlines VREs aid the process of biomedical analysis tracking and the following section introduces description-driven systems and describes how they facilitate traceable designs. In section 4 we outline the use of CRISTAL as a VRE for managing a big data application in supporting the construction of an experiment at CERN's Large Hadron Collider (LHC). To demonstrate its generic capability for supporting data analytics, sections 5 and 6 of this paper contrast this with CRISTAL's use in a VRE supporting analyses of MRI images in the search for biomarkers for the onset of Alzheimer's disease. In the final sections of this paper we investigate the common design approach underpinning these applications and outline conclusions and future work.

2. The Role of Virtual Research Environments in Biomedical Analysis

In any research environment where there are many sets of data, and versions of algorithms operating on those data, it is important to retain a record of who did what, to which sets of data, on which dates, and for what purpose as well as recording the results of the analysis process itself [10]. This information needs to be logged in some research support environment so that analyses can be reproduced or amended and repeated as part of rigorous research processes. A Virtual Research Environment is suitable for that purpose [1]. VRE features usually include collaboration support (Web forums and wikis), document hosting, and some discipline-specific tools, such as data analysis, visualisation, or simulation management. They have become important in fields where research is primarily carried out in teams which span institutions and even countries: the ability to easily share traceable information and research results is central to their use.

VREs should support the entire research lifecycle from the creation of the research ideas and processes through the orchestration of the research processes (or workflows) to the logging of research results and their publication to the academic world. In biomedical analysis reproducibility of results and the sharing of the analyses between researchers is particularly important. There have been a number of software frameworks that have been developed over the past decade to provide the functionality of VREs; some have been applied to biomedical research. The myExperiment VRE was developed to enable collaboration and the sharing of experiments [11] and digital items associated with the research being tracked (so-called research objects). It is based on the Taverna workbench [12] which provides support for scientific workflow execution. Workflows enable the collection of provenance information about research object, to record the enactment of the workflows and to log the workflow outcomes. MyExperiment has been used in bioinformatics applications successfully but it does not allow dynamic redefinition of workflows 'on-the-fly' nor does it provide orchestration of the research process using instantiations of workflow descriptions as would be required for automated control of biomedical analyses. Orchestration of workflow enactment is required when gold standard processes need to be rigidly followed (e.g. in medical analyses) or when analyses are being independently verified.

Another example in medical imaging, is the UK's JISC-sponsored VRE-CI (Virtual Research Environment for Cancer Imaging) [13] project. VRE-CI uses the Microsoft SharePoint software to enabled researchers from multiple disciplines working with different institutions in geographically dispersed locations to collaborate over the development of novel algorithms for image segmentation. VRE-CI allows these researchers and clinicians to share information, images and algorithms and thereby to assess the performance of new developed algorithms and ground truth images. It acts as a project repository for images and algorithms required for the analysis of cancer images but it cannot provide the control and orchestration required to ensure reproducibility of analysis workflows nor is it dynamically reconfigurable. Other projects that have also provided (limited) VRE functionality include the DILIGENT, D4Science and D4Science-II set of EU collaborations [14] and these have been applied to users in biodiversity and humanities research. A full summary of recent VRE projects and a discussion of their achievements and limitations can be found in [1].

The research outlined in this paper provides VRE support for the orchestration of biomedical research based on data and workflow-tracking in an environment that allows dynamic reconfiguration of the process (and data) definitions and thus around-the-clock operation. The approach advocated in the following section is that of capturing the high level description of system objects (and their meta-data) which are then instantiated on an as-needed basis to support the activities of biomedical researchers, providing an active level of support for the research community that enables the reproducibility and verification of results between teams of researchers.

## 3. Description-Driven Systems based Provenance Capture and VREs

In a collaborative research environment, where researchers may use others' methods and results, the traceability of the data that is generated, then stored and subsequently used is therefore clearly important [15]. These forms of knowledge are these days referred to as historical provenance information. The availability of provenance information (the history, evolution and usage for example) in research analytics is often as valuable as the results of the research analysis itself [16]. All of this provenance information, generated by the execution of workflows enables the traceability of the origins of data (and processes) and, more importantly, their evolution between different phases of their usage. Capturing and managing this provenance data in a VRE allows users to query analysis information, to automatically generate workflows and to identify errors and unusual behaviour in previous analyses.

Provenance essentially means the history, ownership and usage of data and its processing in a domain under study. For example, logging the processing of datasets in the study of MRI scans to determine biomarkers of the onset of Alzheimer's disease as in [17]. In this example, the knowledge acquired from executing neuroimaging pipelines (or workflows) must be validated using acquired provenance information. Research has recently been carried out in bio-medical informatics into the provision of infrastructures (whether Grid or Cloud-based) to support researchers for data capture, image analysis, the processing of scientific workflows and the sharing of results. This support includes browsing data sets and specifying and executing workflows (or pipelines) of algorithms required for neurological analytics and the visualization of the results [9]. To date no-one has considered how such analyses can be tracked over time, between researchers and over varying data samples and analytics workflows *as the analysis process itself evolves*. This paper addresses that deficiency by outlining the use of self-describing structures in what we call "Description-driven systems".

Description-driven systems (DDS) design, as we advocate, involves identifying at the outset of the design process all the critical elements (such as business objects, lifecycles, goals, agents and outcomes) in the system being considered and creating high-level descriptions of these elements which are then stored in a model, that can be *dynamically modified* and managed separately from their instances. A DDS, as detailed in [18], makes use of so-called meta-objects to store domain-specific system descriptions, which control and manage the lifecycles of meta-object instances, or domain objects. In a description-driven system, descriptions are managed independently to allow the descriptions to be specified and to evolve asynchronously from specific instantiations of those descriptions. Separating all descriptions in the model from their instantiations allows multiple versions of items to coexist with older versions. This separation is crucially important in handling complexity issues facing many big data applications and allows the realization of interoperability, reusability and system evolution which can underpin a VRE.

The DDS orchestrates the execution of the processes defined in that model (with the consequent capture of provenance information) and handles system evolution *on-the-fly*. Data and process descriptions in the DDS can be modified at runtime and can cater for system updates without user interventions, code recompilation or downtime. This flexibility has been proven both by the development and use of the CRISTAL software over a period of 12 years in ta VRE for supporting the construction and operation of the CMS Electromagnetic calorimeter (ECal) [19] at CERN, as reported in the next section, and its application to the medical data domain, described later in this paper.

## 4. CRISTAL as developed at CERN and as the basis for Medical Data Analytics

To demonstrate the benefits of adopting a description–driven design approach to developing a platform for data analytics consider the environment in which our CRISTAL system was developed at CERN. Physicists at CERN construct and operate complex experimental equipment whose construction processes are highly distributed, very data-intensive and require a computer-based VRE that tracks the construction processes. In constructing detectors like the Compact Muon Solenoid [19] scientists require management systems that can cope with high levels of data complexity, with systems that can evolve repeatedly over time (usually because of extended development timescales and consequently changes in user requirements) and with scalability in terms of data volume. They require very fine granularity of provenance information over long timescales to facilitate longitudinal data analytics. In essence they require the functionality of a VRE to provide a platform for information capture and sharing.

The CMS construction required detector parts of different model versions to be handled over the complete usage lifecycles and for these model versions to coexist and be stored alongside different model versions. In many ways this is very similar to biomedical informatics (and many other scientific disciplines) where data sets are model-based, long-lived and are accessed by evolving suites of algorithms and by groups working collaboratively, often geographically distributed. By separating details of model types from the details of instantiated data elements in CRISTAL we enabled the model type versions to be specified and versioned independently and explicitly from single data elements (i.e. the instantiations of the types). In capturing descriptions separately from their instantiations, system evolution can be handled while analysis was underway and continuity in the analysis process could be supported. CRISTAL can be considered a distributed product, document, meta-data and workflow management system which makes use of a multi-layered architecture for its component abstraction and dynamic object modelling (see figure 1) for the design of the objects and components of the system that is being designed [20]. A full description of CRISTAL can be found in [3] which details the history of DDS development. In short it provides versioning for data and processes (and their descriptions) allowing multiple versions of item instances and descriptions to co-exist and keeps track of how activities have been applied to items in analyses (provenance). This essentially makes the CRISTAL VRE self-describing; there is a separation of concerns between the model design and any application software that accesses the model.

As noted in following a DDS approach the separation of object from object description instances was required. This abstraction resulted in a three layer description-driven architecture being developed (as shown in figure 1). Our CRISTAL approach is similar to the model-driven design concept espoused by the OMG [21], but differs in that the descriptions and the instances of those descriptions are implemented as objects (Items); crucially they are implemented and maintained using exactly the same internal model, promoting conceptual simplicity and ease of system maintenance. Even though workflow descriptions and data implementations are different they are both saved as Items in CRISTAL, and, as a consequence, the way in which they are stored and are related to each other is exactly the same. Our design approach is somewhat similar to the distinction between Classes and Objects in the definition of object oriented principles [22]. We have followed those principles in CRISTAL so that we can provide the required level of system maintainability, reusability and flexibility to facilitate evolution and the gathering of longitudinal provenance information. Within CRISTAL each defined system element (or Item) is stored and versioned. This allows users of the system to view older versions of their Items at a later date and either extend a version of an Item or return to an earlier version of an Item. This functionality underpins the use of CRISTAL-resident data for data analytics (as described later in section 5 of this paper) and as the basis for a VRE. In CRISTAL a model is designed as a set of Items, each having a lifecycle defined by workflows, themselves composed of activities. The activities comprising an Item's lifecycle have similarities with object oriented methods, since they define a single action performed on that Item. Each activity set must always form a valid graph of activities from the creation of the Item to its completion. This clarity of design through implementation constraints is a return to the philosophy advocated in early OO languages such as Smalltalk [23], which steered towards a core design with the system logic partitioned in a manageable way.

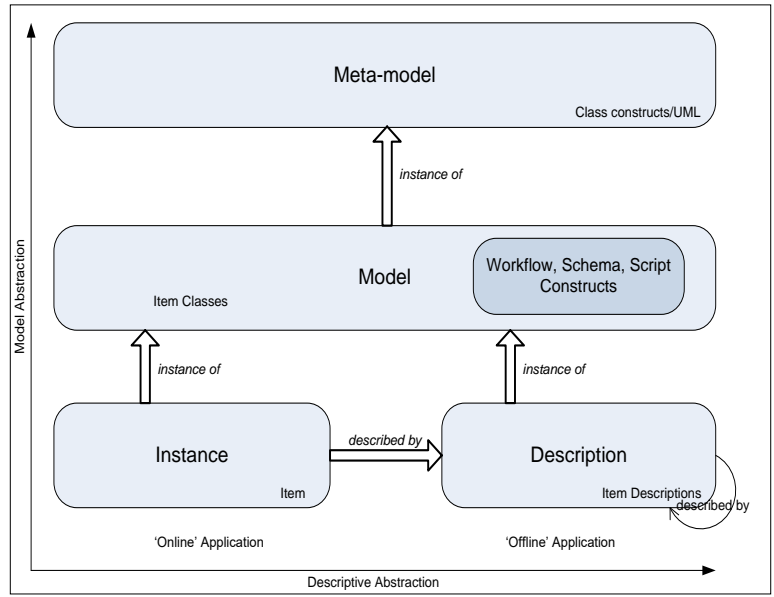

Figure 1: Model versus Description in CRISTAL (from [3])

The application logic that needs to be executed during a workflow has its functionality defined along with its constituent activities. It is then simple to import these definitions into the system under consideration where the functionality can be evaluated for feedback to the users. All this can be done without recompiling a single line of code or restarting the application server. As a consequence application users can manage their Item definitions, can have their activities orchestrated and have appropriate provenance data (on Item and system use and outcomes and results) captured that facilitates their data analytics. In our experience, the process of factoring the lifecycle and dataset of the new item type into activities and their outcomes helps understanding of the desired functionality in the user's mind. In practice this has been verified over a period of more than a decade of use of CRISTAL at CERN.

## 5. Medical Analytics and Traceability using CRISTAL in a VRE

Given its successful record for supporting data analytics at CERN, CRISTAL was selected as the basis of a VRE to support medical image analyses in the EC Framework 7 projects neuGRID and neuGRID for Users (N4U) in studies of Alzheimer's disease. The full details of these studies are beyond the scope of the current paper (details can be found in [24]); they serve to illustrate the functions of a DDS as used for tracing scientific workflows and supporting biomedical data analytics in a VRE. In the N4U project we have delivered a Virtual Laboratory (VL [25]) or VRE as the platform for their analyses offering neuroscientists tracked access to a wide range of Grid- or Cloud-resident big data sets and services, and support for their studies of biomarkers for identifying the onset of Alzheimer's disease. The N4U VL, is based on services layered on top of the neuGRID infrastructure and a CRISTAL database and is shown diagrammatically in figure 2. The Virtual Laboratory has been designed to be reusable across biomedical research communities. The VL enables researchers to conduct their studies and analyses by locating longitudinal clinical data, pipelines, algorithm applications, analysis definitions and detailed provenance data in a user-accessible environment. This has been achieved by basing the N4U VL on a so-called integrated VRE Analysis Base (or Data Atlas, as outlined in the sister paper to this paper [26]) which gathers together all provenance-enabled objects (datasets, workflows/pipelines, user analyses and their meta-data), into a single knowledge base for access by information, querying, visualization and analysis services. This Analysis Base is directly accessible by a set of Information Services and Analysis Services; these three elements are the building blocks of a VRE.

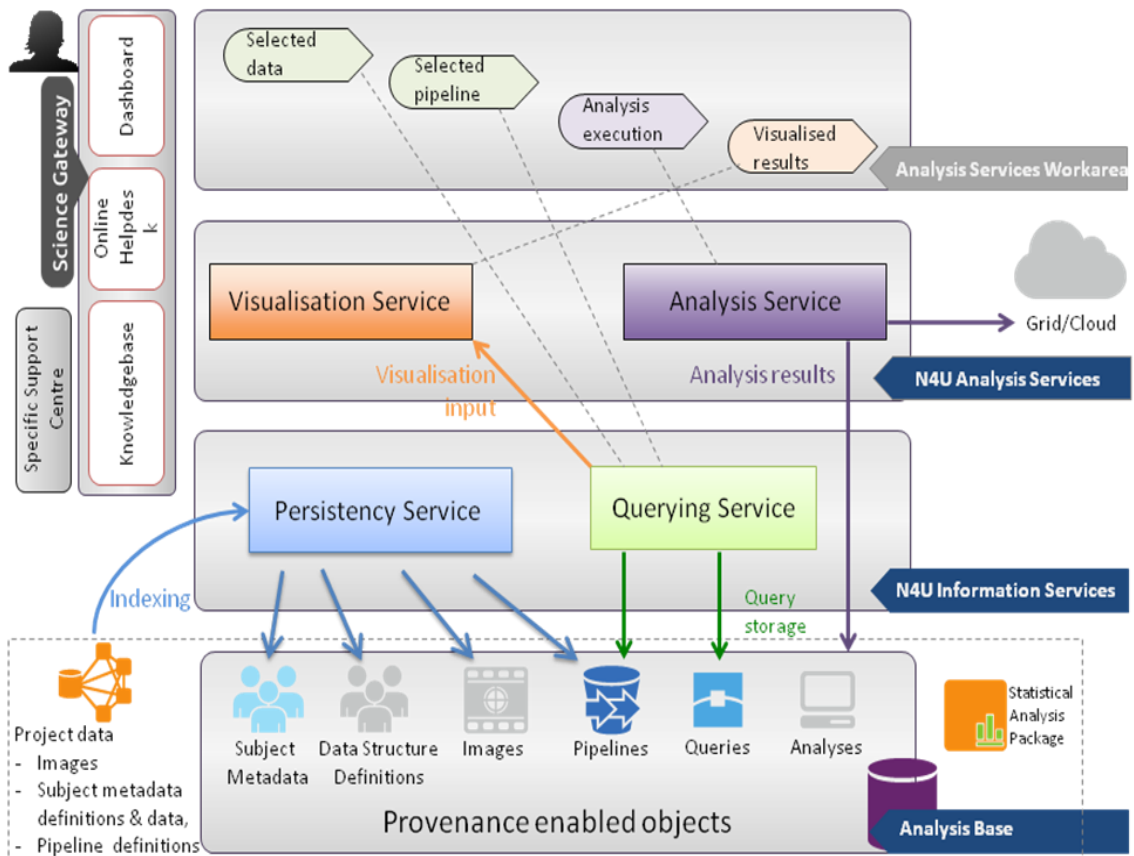

Figure 2 : The neuGRIDforUsers (N4U) Virtual Laboratory / VRE

The high-level flow of data and analysis operations between various components of the VL and the Analysis Base are also highlighted in figure 2. The N4U Analysis Base enables analytics by indexing and interlinking the clinical study datasets stored on the N4U Grid infrastructure, their algorithms and scientific workflow definitions and their resultant provenance information. Once researchers conduct their analyses using this information, the analysis definitions and resulting data are made available in the VRE Analysis Base for tracking and reusability purposes via an Analysis Service. In N4U, we used CRISTAL to capture the provenance data needed to support neuroimaging analyses and to track heterogeneous datasets and their associated metadata, individualized analysis definitions, and usage patterns, thus creating for all intents and purposes a VRE knowledge base for the community of researchers. The VRE Analysis Base stores datasets, clinical study data, meta-data, queries, pipelines and outputs of analyses. It is accessed through services that store or index datasets and algorithm definitions so that users can perform any investigation or data analysis using the VL (see section 5). This subset of Information Services is termed the Persistency Service.

The Information Services also provide interfaces for the parameterised querying of datasets to assist VL users in defining and executing their analyses on datasets. This subset of Information Services is termed the Querying Service. The outcome generated by the Querying Service can be exported in various formats, such as XML and CSV, which is then used in other software applications to generate or to perform an analysis, e.g. by using the CRISTAL software [3]. The Persistency Service indexes source data and pipelines for use in biomedical analyses. These come from a wide range of external sources which may be existing data sets that we are indexing and integrating into the N4U environment or they may be new data sets as a result of analyses. This results in a large amount of structural heterogeneity between datasets. Using CRISTAL's VRE platform, the Persistency Service uses a meta-schema approach to store this data.

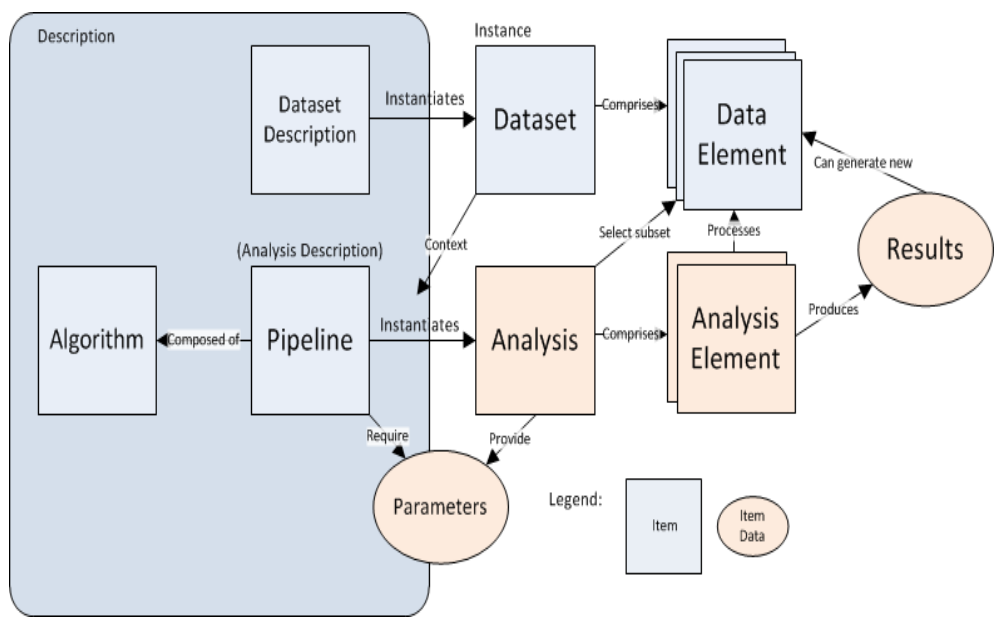

Figure 3 : Analysis Service Architecture.

The Analysis Service is the core of the entire VL/VRE; it uses CRISTAL as its backend and provides access to the tracked information for browsing, visualization, pipeline authoring and execution. The Analysis Service is responsible for executing jobs on the N4U infrastructure as well as for provenance capture. It captures provenance data that emerges in the specification and execution of the stages in analysis pipelines. It also keeps track of the origins of the data products generated during analysis and any evolution between different stages of that analysis; it records every change made to its Items (or objects) as shown in figure 2. Whenever a change is made to any Item, the definition of that unit of data or any application logic applied to that data, it stores the modification and the metadata associated with that modification (e.g. who made the change, when, where, how and for what purpose) alongside the Item. This makes CRISTAL applications fully traceable, and this data may be used to gather together the detailed provenance information for subsequent examination. In N4U, the Querying Service (figure 2) provides provenance querying facilities for the VL/VRE.

Part of the CRISTAL VRE model is shown in figure 3; it shows CRISTAL Items as squares and external project data portrayed as ovals. Objects associated with user analyses are shown in orange. The model indicates how pipelines (workflows) and datasets are used to create "parametrized" analyses that yield results as outcomes. The figure also shows objects called Data Elements and Analysis Elements which provide the substance of all the Provenance Enabled Objects stored in the VRE Analysis Base (as shown in figure 2). The model contains Items holding metadata on all the pipelines and datasets registered in N4U. Pipeline Items give the location of the analysis scripts that need to be run, along with default execution environment settings, and any common directory locations that should be passed with the job. Dataset Items comprise Data Elements, which are sets of files that should be processed together in one single job, along with the specific metadata about that set that can vary in composition between different datasets [24].

The central Item type of the Analysis Service itself is the so-called 'Analysis' object (figure 3), which is a user-initiated process orchestrating the execution of a selected pipeline on one or more elements of a single selected dataset. Each Analysis belongs to the user that created it, and can only be viewed by that user, this information also being logged on execution by the CRISTAL VRE. In execution the Analysis suite instantiates an Analysis Element for each given Data Element, creating an instance of the Pipeline workflow that can be dispatched to the Grid (or Cloud) as executable jobs and whose provenance is gathered for each step of the workflow execution to be tracked by CRISTAL. The Analysis Service provides workflow orchestration for scientists and a platform for them to execute their experiments on the Grid (or Cloud). It allows users to recreate experiments on the neuGRID Infrastructure (for comparison, verification, reproducibility) using previously recorded provenance.

The Analysis Service enables data analytics by controlling:

- The browsing of previous analyses and their results sets;
- The specification of new analyses by pairing selected datasets with selected algorithms and pipelines found in the VRE Analysis Base;
- The execution of analyses by creating jobs to be passed to the N4U Pipeline Service then, once the jobs have executed, logging the returned results in the analyses objects;
- Any re-running of past analyses with different parameters or altered datasets and
- Enabling the sharing of analyses between researchers.

Once the specification of an analysis is complete it can be run. This involves sending jobs out to the Grid (or Cloud) for every element of the analysis algorithm selected, for every element of the selected dataset. This could potentially generate a large amount of parallel work, which the Grid (or Cloud) will then distribute to many computing elements. As the elements of the child analysis complete, the Analysis Service keeps a record of the state of the result set, which Grid/Cloud resources were used for the computation, and the exact times the operation was started and completed as part of the provenance data for that analysis. All of the data resulting from the analysis and its provenance is stored in CRISTAL and is queryable via the Querying Service. All analysis objects belong to, and are only visible to, the user who created them; it is possible to share analyses between users in the project.

## 6. Evaluation of the VRE in Practice

The operation of the CRISTAL VRE is best understood via a practical example. Consider our medical imaging example where a clinician wishes to carry out a new analysis. The stages of that analysis are likely to be i) Investigation; ii) Definition; iii) Execution and iv) Consolidation. In the Investigation phase she may want to browse earlier analyses to help clarify the subject matter of her analysis. She may want to find out what other researchers' analyses have been conducted, what were their findings, what data and/or workflows they had used for their studies etc. To do this she would browse the VRE Analysis Base using the Querying Service to look at how previous analyses were defined, who, what, and when they had been carried out, what were their finding and what data processing had been carried out. On the basis of this she may amend and existing workflow/dataset/analysis process or define an entirely new analysis.

Once her investigations were complete she would enter the Definition phase in which she would make a selection of data from the datasets which are available to her. For this she would log into the Analysis Service Area and interact with the Querying Service to identify the dataset that possesses the particular properties she needs. In N4U she would access the system through the 'Science Gateway' dashboard of its VRE to invoke the Querying Service, as shown in figure 2. She would then submit her constraints, which are passed as predicates to a query for handling by the Querying Service. The Querying Service then queries the VRE Analysis Base and returns a list of valid dataset properties and locations which meet her constraints. The Querying Service would then display this list to the clinician fro approval and she can then declare this as her analysis '*working dataset*'. Once she is satisfied with her dataset selection she would then use the Querying Service to identify a pipeline (or workflow) specification to create her overall analysis. To do this she would need to use the Analysis Service Interface to search CRISTAL for existing algorithms that she can use to create a new pipeline or to select a pre-defined pipeline. Her analysis is defined as an instantiation of a pipeline in the context of a dataset and a pipeline. The completed pipeline (her analysis '*working pipeline*') will have her analysis 'working dataset' associated.

In the Execution phase her analysis will be run on each element of the analysis working dataset by CRISTAL, as described in section 4. The interaction is performed through a portlet base UI (for novice users) or from the command line for advanced users. A web service endpoint is also available for users that wish to write more complex programs. The working pipeline would then be sent to CRISTAL which would orchestrate the stages in the pipeline using the N4U GridBroker and its Pipeline Service (figure 4). This means that a single activity from the analysis working pipeline would be sent to the Pipeline Service as a single job. Once the job has completed, the result would be returned to CRISTAL and it would extract and store provenance information for this job. This information would contain traceability factors such as the time taken for execution, and whether the job completed successfully. It would store this information internally in its own data model. It would also post this information to the VRE Analysis Base so that this important provenance information would be accessible by the Querying Service, potentially by other users at a later date.

This loop of sending jobs for each activity in the workflow and receiving the result would continue until the workflow is complete. Once this workflow has completed CRISTAL would once more generate provenance information and store this for the entire workflow in its own internal data store and the VRE Analysis Base. In the Consolidation phase the final result of the completed pipeline would be presented to the user for evaluation. She would invoke the Visualization Service to examine the outputs of her analysis and, where appropriate, she would annotate the outcomes again to potentially be queried by other users at a later date. The clinician now has a permanently logged record (including provenance) of her analysis including the datasets and algorithms she has invoked, the data captured during the execution of her analysis and the final outcome returned by her analyses. These provenance elements may also have associated annotation that she has added to provide further knowledge of her analysis so others could consult at a later time to re-run, refine or verify the analysis that has been carried out.

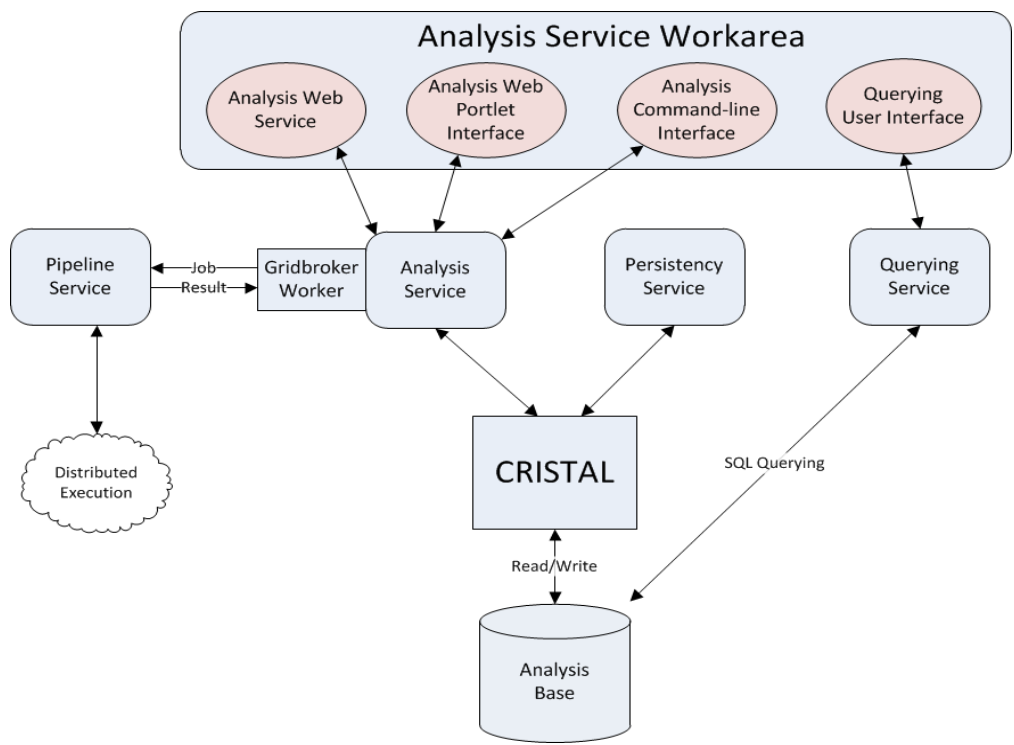

Figure 4 : Detail of the CRISTAL VRE Model

The examples that have been described above of the CRISTAL VRE usage across the spectrum of information systems from CERN's High Energy Physics studies to Medical Imaging show the flexibility of the description-driven design approach to systems design and implementation. They also demonstrate the role of provenance data collection, its management and its application in the traceability of data and workflows for data analytics. These techniques are actually widely applicable to any big data development for (Grid or) Cloud-resident data. If designers follow such a description-driven approach they can concentrate on the important building blocks of their systems - the business-critical enterprise objects whose lifecycles (creation, versioning, usage, evolution and final deletion) require tracing for the business to function properly. In many ways this is similar to the emerging field of 'domain driven design' described fully in [27]. CRISTAL 'Items' may be any of data elements or sets, images, composite or atomic activities, agents, roles or people. As in domain-driven design these objects are specified at the outset of the design process, in our case in an instance of CRISTAL that supports a flexible, described and extensible data model. However in contrast to domain-driven design CRISTAL enables full lifecycle traceability and orchestration of those Items in system usage and thus provides a complete behavioural picture of the system from creation to completion.

Once designers have identified the Items of interest in their system the Items are assigned descriptions and meta-data is then associated with both the Items and their descriptions and the designer considers questions such as: when are the Items created and by whom and for what purpose? Who can change them over time, how, when and why? What data does each Item generate as outcomes or results when activities/algorithms are run against those Items? By what mechanisms can the Item be viewed and for what purpose? Does the Item persist over time and in which versions can it be used concurrently? These forms of provenance data are often referred to as the sevens "W"s: who, which, what, when, where, why and hoW [28]. Note here we refer to the Item and its use across the business model rather than seeing the business model as supporting a specific application (for example a process based workflow management system). Thus the very same 'Person' Item (and its description and meta-data) can be viewed via a Project Management system, a Human Resources system, a Payroll system, a Project Management system, a Resource Scheduling system or whatever functions that are critical to the operation of that enterprise. Its specification and access control is not dependent on its ultimate implementation environment.

## 7. Conclusions

The studies described in this paper have shown that describing a biomedical system explicitly and openly in a model (in this case a DDS model) as part of a VRE enables software engineers to evolve aspects of it incrementally as the system requirements change. This enables system transition from one version to the next with uninterrupted system availability and facilitates traceability of data and processes throughout the system's lifecycle. In a practical sense this leads to a high level of control over the evolution of the system design and allows easier implementation of system improvements and simpler maintenance over time. Future work is ongoing to model domain-related semantics, for example the specifics of a particular application domain such as in medicine, biomedical analytis, health, transport or other domains. Over time this will allow CRISTAL to develop into a fully self-describing model execution and orchestration engine, making it possible to build applications directly on top of the design, without significant new code generation. This will allow usage patterns to be captured, roles and agents to be defined on a per-application basis, and rules specific to particular domains to be captured and curated, all in the CRISTAL-enabled VRE. In addition such information will enable multiple instances of CRISTAL to discover the semantics required to inter-operate and to exchange data. To support this in Q2 2016 a version of CRISTAL called CRISTAL-ISE was released to the public as Open Source under the LGPL V3.0 licensing scheme (see http://cristal-ise.github.io/). It can be used as the basis of a VRE not only in scientific applications, but also in business-oriented applications as demonstrated by its use in supporting Business Process Management in the CRISTAL-ISE project by the French company M1i [29].

In the long term we will further develop a so-called Provenance Behaviour module for CRISTAL. This will enable its functionality as a (bio-)medical VRE to be extended with domain-specifics such as user profiles, 'my-analysis' workflows and user-specified data visualizations. Using machine learning approaches we aim to formulate models that can derive optimisation strategies by learning from the past execution of processes. These models would evolve over time and would facilitate decision support in the design, implementation and execution of future workflows in associated domains. CRISTAL could then employ approaches to learn from the data that has been produced, find common patterns and models, classify and reason from the information accumulated and present it to the system in an intuitive way. This information will be delivered to users as acquired knowledge of system behaviour and would guide work on new workflows and underpin future design decision-making.

One essential future element is the provenance interoperability aspect present within the neuGRID/N4U projects. Currently, we are exporting the provenance enabled objects to the emerging PROV [30] interoperability standard. This will allow N4U users to use their provenance data in other PROV-compliant systems. There are plans to enrich the CRISTAL kernel (the data model) to model not only data and processes (products and activities as Items) but also to model agents and users of the system (whether human or computational). We will investigate how the semantics of CRISTAL items and agents could be captured in terms of ontologies and thus mapped onto or merged with existing ontologies for the benefit of new domain models.

**Acknowledgments**

The author wishes to highlight the support of their home institutes and acknowledge funding from the European Union Seventh Framework Programme (FP7/2007-2013) under grant agreement n. 211714 ("neuGRID") and n. 283562 ("neuGRID for users") with special thanks to the N4U consortium.